\documentclass{osa-article}

\journal{oe}
\begin{document}

\title{Efficient Wave Optics Modeling of \\Nanowire Solar Cells Using \\Rigorous Coupled Wave Analysis}

\author{Kyle W. Robertson,\authormark{1} Ray R. LaPierre,\authormark{2} and \\
Jacob J. Krich\authormark{1,3}}

\address{\authormark{1}Department of Physics, University of Ottawa, Ottawa, ON, K1N 6N5, Canada\\
\authormark{2}Department of Engineering Physics, McMaster University, Hamilton, Ontario, L8S 4L7, Canada\\
\authormark{3}School of Electrical Engineering and Computer Science, University of Ottawa, Ottawa, Ontario, K1N 6N5, Canada}

\email{\authormark{1}krobe088@uottawa.ca} %% email address is required

% \homepage{http:...} %% author's URL, if desired

%%%%%%%%%%%%%%%%%%% abstract %%%%%%%%%%%%%%%%
%% [use \begin{abstract*}...\end{abstract*} if exempt from copyright]

\begin{abstract}
We investigate the accuracy of rigorous coupled wave analysis (RCWA) for
near-field computations within cylindrical GaAs nanowire solar cells and
discover excellent accuracy with low computational cost at long incident
wavelengths, but poor accuracy at short incident wavelengths. These near fields
give the carrier generation rate, and their accurate determination is essential
for device modeling. We implement two techniques for increasing the accuracy of
the near fields generated by RCWA, and give some guidance on parameters
required for convergence along with an estimate of their associated computation
times. The first improvement removes Gibbs phenomenon artifacts from the
RCWA fields, and the second uses the extremely well-converged far field
absorption to rescale the local fields.  These improvements allow a
computational speedup between 30 and 1000 times for spectrally integrated
calculations, depending on the density of the near fields desired.  Some
spectrally resolved quantities, especially at short wavelengths, remain
expensive, but RCWA is still an excellent method for performing those calculations.
These improvements open up the possibility of using RCWA for low cost optical
modeling in a full optoelectronic device model of nanowire solar cells.
\end{abstract}

%%%%%%%%%%%%%%%%%%%%%%%%%%  body  %%%%%%%%%%%%%%%%%%%%%%%%%%

\bibliography{main}

\section{Introduction}

Nanowire solar cells (NWSC) are a new solar cell technology with the potential to
improve upon existing solar cell devices. Their potential stems
from their ability to effectively absorb incident light while using less
semiconductor material than planar solar cells.

The optimal nanowire solar cell arrays consists of nanowires that are a few
microns in height and with diameters and periodicities comparable to the
wavelengths present in the solar spectrum \cite{wu_analytic_2017,
sturmberg_optimizing_2014}.  These small sizes require full wave optics
simulations to accurately model their optical properties, unlike in standard
planar devices \cite{wallentin_inp_2013, kupec_light_2010}.  Both experimental
measurements and modeling have shown high levels of absorption with low
sensitivity to the incident angle of light \cite{ghahfarokhi_performance_2016}.
Additionally, the finite in-plane dimensions of nanowires can accommodate
strain due to growth on lattice-mismatched substrates without introducing
dislocation faults in the crystal lattice \cite{kavanagh_misfit_2010}. This
capability opens up the possibility for III-V tandem cells grown on silicon
\cite{borgstrom_towards_2018}.

The larger design parameter space of NWSC relative to planar solar cells
requires careful optimization of geometric parameters to maximize device
performance \cite{robertson_optical_2017}. There is a need for fast, accurate
modeling tools to enable rapid exploration and optimization of nanowire
designs. Conventionally, finite element \cite{hu_optical_2012, kupec_light_2010,
azizur-rahman_wavelength-selective_2015,trojnar_optimizations_2016} and
finite difference methods \cite{wu_analytic_2017,fountaine_resonant_2014,
du_broadband_2011} have been used in optical models of NWSC. While these
techniques are highly accurate, they are computationally expensive,
limiting their usefulness in a closed-loop global device optimization.
Rigorous coupled wave analysis (RCWA) is another wave-optics modeling
technique that lacks the memory and computational requirements of competing
techniques \cite{moharam_formulation_1995}. RCWA is a Fourier domain
technique ideally suited to periodic arrays. It is promising for its speed
and is highly accurate when computing far-field quantities such as total
absorptance, reflectance, and transmittance. RCWA simulations become more
accurate as the number of plane waves $N_G$ increases, and the computational
cost scales as $N_G^3$. However, naive implementations
lack accuracy at reasonable $N_G$ when computing near-fields internal to the device due to the
well-known Gibbs phenomenon \cite{weismann_accurate_2015}.  Such
near-fields are required to compute carrier generation rates and are thus
an essential component of a fully-coupled optoelectronic device model.

\begin{figure}[htpb]
    \centering
    \includegraphics[width=0.8\linewidth]{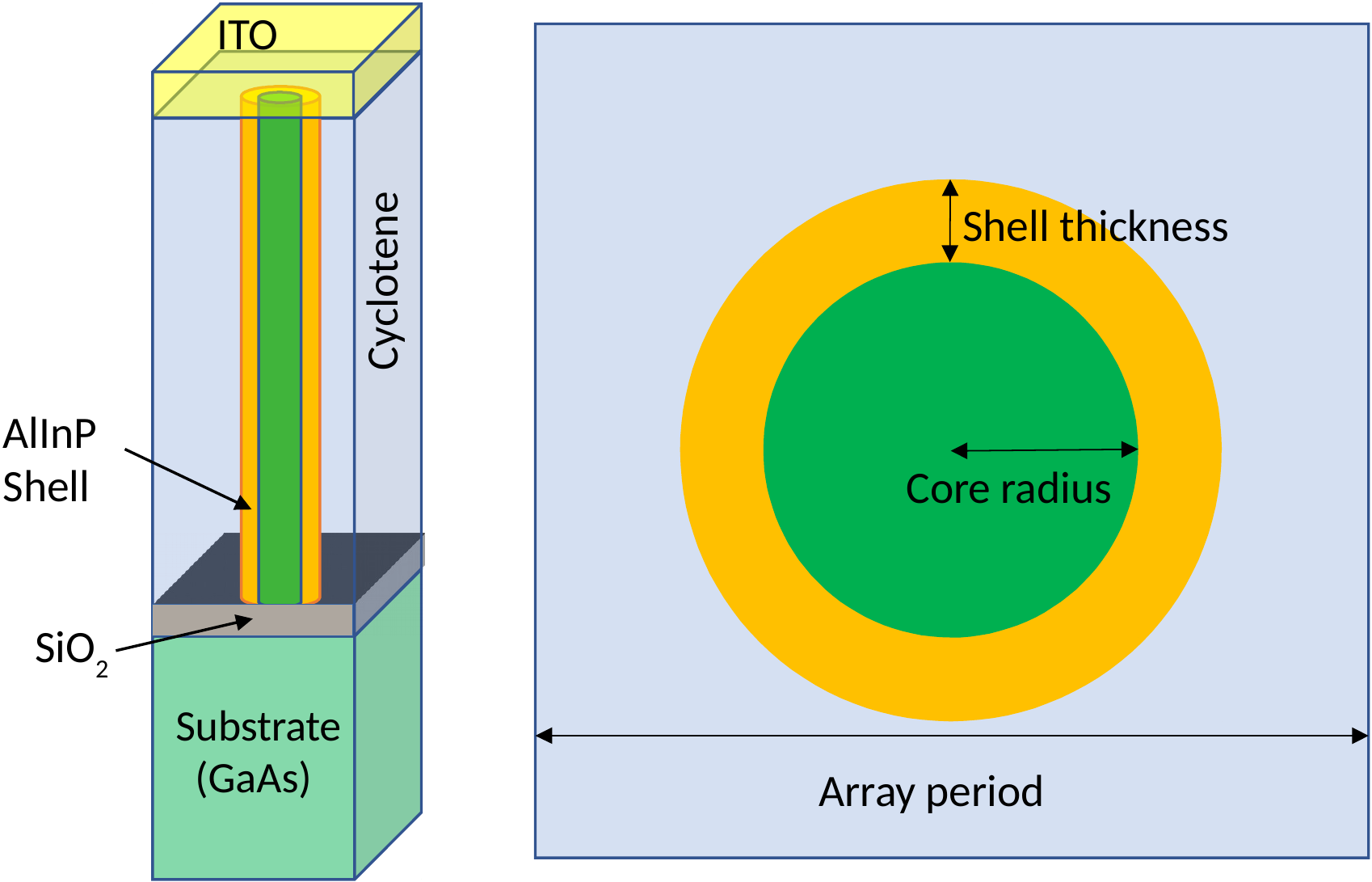}
    \caption{The test device used for assessment of RCWA. Left: A single unit
        cell in a square nanowire array containing a cylindrical GasAs nanowire
        passivated by an AlInP shell on a GaAs substrate, planarized by a
        cyclotene dielectric, and top-contacted with a layer of indium tin
        oxide. A thin layer of SiO$_2$, surrounding the GaAs core but lacking
        the AlInP shell, exists between the cyclotene and substrate. Right: A
        top down view of the unit cell, demonstrating the piecewise constant
        material parameters in the plane.} 
    \label{fig:nw-schematic} 
\end{figure}

In this work, we assess the accuracy of RCWA for use in optical modeling of
nanowire solar cells. We examine a test device (see Fig.~\ref{fig:nw-schematic}
and Table \ref{tbl:nw_params}), indicate where RCWA lacks a desirable level of
accuracy, and provide two techniques for increasing accuracy of the near
fields. The first is an implementation of an already published technique for
introducing proper discontinuities in the near fields and mitigating the Gibbs
phenomenon
\cite{lalanne_computation_1998,brenner_aspects_2010,weismann_accurate_2015,bezus_stable_2012}.
The second is a new rescaling technique that increases the accuracy of device
simulations while keeping computational cost reasonable when computing
spectrally integrated quantities.  We show that even with our two improvements,
some spectrally resolved quantities continue to require more expensive
calculations. Using our improvements, RCWA shows promise as an effective
technique for rapid optical modeling of nanowire solar cells.

\begin{table}[]
    \centering
    \begin{tabular}{|l|l|}
    \hline
    \textbf{Parameter}  & \textbf{Value} \\ \hline
    NW Core Length      & 1.3 $\mu \mathrm{m}$    \\
    NW Shell Length     & 1.27 $\mu \mathrm{m}$    \\
    SiO$_2$ Thickness   & 30 nm       \\
    Substrate Thickness & 1 $\mu \mathrm{m}$       \\
    ITO Thickness       & 300 nm         \\
    Array Period        & 250 nm         \\
    Core Radius         & 60 nm          \\
    Shell Thickness     & 20 nm          \\ \hline
    \end{tabular}
    \caption{Numerical values for all geometric parameters in the test device.}
    \label{tbl:nw_params}
\end{table}

\section{RCWA}

RCWA is a Fourier-space method for solving the source-free frequency domain Maxwell's
equations:

\begin{align}
    \nabla \times \mathbf{H} &= - i \omega \epsilon \mathbf{E} \label{eq:maxwells-eqs} \\
    \nabla \times \mathbf{E} &= i \omega \mu \mathbf{H} \\
    \nabla \cdot \mathbf{H} &= 0\\
    \nabla \cdot \mathbf{E} &=0
\end{align}
where $\mathbf{H}$ is the magnetic field, $\mathbf{E}$ is the electric field,
$\omega$ is the oscillation frequency, $\epsilon$ is the electric
permitivitty, and $\mu$ is the magnetic permeability. RCWA relies on two
critical assumptions about the geometry of the system.  First, the device must
be composed of discrete, axially-invariant layers such that at a given $x$-$y$
point within a layer, the material parameters along the z-direction remain
constant. Second, the device must be decomposable into fundamental unit cells
that are 2D periodic in the plane.  If these conditions are
satisfied, then the longitudinal and transverse dimensions are separable and
the fields in a single layer can be written as:

\begin{equation}
    \mathbf{H}(\mathbf{r}, z) = \sum_{\mathbf{G}}
    \mathbf{H}_{\mathbf{G}}(z)e^{i(\mathbf{k} + \mathbf{G}) \cdot \mathbf{r}},
    \label{eq:basic-fields}
\end{equation}
where $\mathbf{G}$ is one of $N_G$ in-plane reciprocal lattice vectors,
$\mathbf{k}$ is the in-plane component of the excitation, and $\mathbf{r} = x
\mathbf{\hat{x}} + y \mathbf{\hat{y}}$. Note the ${\mathbf{G}}$ is
generally chosen to be an array of reciprocal lattice points with a
circular truncation, keeping all $\mathbf{G}$ with $|\mathbf{G}|$ less than
some constant, which maintains symmetry in Fourier space \cite{liu_s4_2012}.
The in-plane dielectric profile $\epsilon(\mathbf{r})$ may depend on the
material, allowing it to have piecewise-constant dependence on the transverse
spatial coordinates.  Vertical nanowire arrays (see
Fig.~\ref{fig:nw-schematic}) satisfy these geometric constraints.

The essential part of RCWA is determining $\mathbf{H}_{\mathbf{G}}(z)$ in
Eq.~\eqref{eq:basic-fields} for a given set of reciprocal lattice vectors
$\mathbf{G}$. One can assume the coefficients in Eq.~\eqref{eq:basic-fields}
take the form \cite{liu_s4_2012}

\begin{equation}
    \mathbf{H}_{\mathbf{G}}(z) = \left [ \phi_{\mathbf{G}, x} \mathbf{\hat{x}} + \phi_{\mathbf{G}, y}
    \mathbf{\hat{y}} - \frac{(k_x + G_x)\phi_{\mathbf{G}, x} + (k_y +
G_y)\phi_{\mathbf{G}, y}}{q} \mathbf{\hat{z}} \right ] e^{iqz} ,
    \label{eq:expansion-coeffs}
\end{equation}
where the $\phi$ are expansion coefficients and the z-component has been chosen
to satisfy the $\nabla \cdot \mathbf{H} = 0$ condition. This form of the fields
illustrates one of the key advantages of RCWA over competing techniques, namely
the analytic dependence on the z coordinate. By inserting Eq.
\eqref{eq:expansion-coeffs} into Eq.~\eqref{eq:maxwells-eqs}, one arrives at an
eigenvalue equation for determining the set of eigenvalues $q$ and the
components of the eigenvectors $\phi$ for a single layer. Once the eigenmodes
of each layer have been determined, multilayer structures are joined together
by introducing propagation amplitudes for the eigenmodes and using the
scattering matrix method to join solutions at layer interfaces
\cite{li_formulation_1996,whittaker_scattering-matrix_1999,
moharam_rigorous_2004,kim_extended_2007,rumpf_improved_2011}. Results increase
in accuracy with $N_G$.  Our work is an extension to S$^4$, an open-source
implementation of RCWA built on the scattering matrix method
\cite{liu_s4_2012}. In the remainder of the manuscript, we refer to S$^4$ as
the standard RCWA method, but it has included a significant number of
improvements from the original RCWA methods; for details, see Ref.\
\citenum{liu_s4_2012}.

For optoelectronic device modeling, we are most concerned with determining the
local carrier generation rate, which is determined from the local electric
field strength in each material. RCWA expresses the fields using the Fourier
series in Eq.~\eqref{eq:basic-fields}. Any finite Fourier series representation
is always continuous, even across in-plane material interfaces, as between the
core and shell of a nanowire. In an exact solution, the normal components of
$\mathbf{E}$ should be discontinuous across material boundaries, but a Fourier
reconstruction requires an intractable number of terms to accurately model such
a discontinuity, even though far-field quantities such as the total absorptance may be well
converged. For any finite $N_G$, standard RCWA-produced fields have spurious
oscillations, especially near material interfaces.

To assess the convergence of RCWA with  $N_G$, we define two methods for 
computing the absorptance of a layer of
the device. The first method relies only on the power exiting from the top of
the layer and the power transmitting through the bottom of the layer. These
powers can be computed entirely in Fourier space \cite{liu_s4_2012}, and
do not suffer from convergence issues in the reconstruction of the near
fields. These emitted powers of layer $i$ are defined as:

\begin{align}
    P_\text{up}^i(\omega) &= \int_\text{top} S_{z}(\omega) dA \label{eq:power_reflected} \\
    P_\text{down}^i(\omega) &= \int_\text{bottom} S_{z}(\omega) dA ,
\end{align}
where $S_z$ is the z-component of the Poynting vector and the integration is
over the top or bottom surface of the unit cell, with the appropriate sign for
emitted power. Considering the top (layer 1) and bottom (layer $n$) together,
the total reflectance and transmittance are

\begin{align}
    R(\omega) &= \frac{P_\text{up}^1(\omega)}{P_\text{in}(\omega)} \\
    T(\omega) &= \frac{P_\text{down}^n(\omega)}{P_\text{in}(\omega)} ,
\end{align}
where $P_{in}$ is the input power of the incident plane wave. 
Then total absorptance is:

\begin{equation}
    A_\text{far field}(\omega) = 1 - R(\omega) - T(\omega).
    \label{eq:absorptance-flux}
\end{equation}
The contribution of a single layer to the device absorptance can be calculated
similarly. We consider the test structure detailed in Table \ref{tbl:nw_params}
and use S$^4$ \cite{liu_s4_2012} to perform RCWA calculations with
normally-incident circularly polarized light. We consider 60 equally spaced
frequencies corresponding to wavelengths from 300 nm to 900 nm, just beyond the
GaAs absorption edge of 871 nm.
\begin{figure}[htpb]
    \centering
    \includegraphics[width=0.8\linewidth]{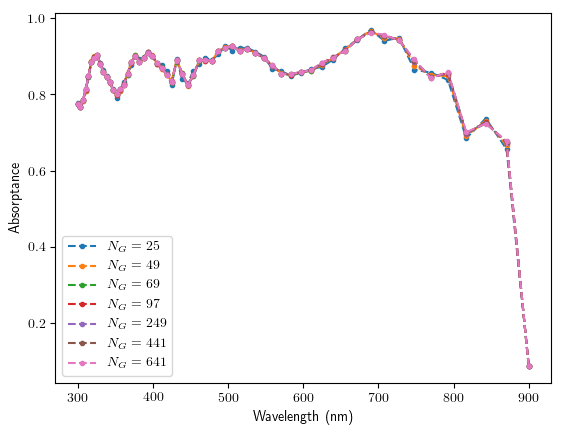}
    \caption{Absorptance of the entire device calculated using the far field
        fluxes,  Eqs.~\eqref{eq:power_reflected} - \eqref{eq:absorptance-flux}.
        The markers for all values of $N_G$ lie nearly on top of one another,
        indicating convergence at low numbers of basis terms.}
    \label{fig:absorptance-farfield} 
\end{figure}
\noindent Figure \ref{fig:absorptance-farfield} shows that the far-field absorptance
spectrum of the full device converges rapidly with basis terms, and is self-converged within 0.5\%
with $N_G = 75$. 

Equation \eqref{eq:absorptance-flux} expresses the power absorbed in a layer in
terms of the fluxes into and out of the layer. The divergence theorem and
Maxwell's equations allow rewriting that power in terms of the local fields,
instead. The absorbed power can then be written,

\begin{align}
    \mathcal{P}_\text{abs}(\omega) &= \epsilon_0 \omega \int n(x,y,z;\omega)
    k(x,y,z;\omega) \lvert E(x,y,z;\omega) \rvert ^2 dV \label{eq:power_absorbed_nearfield} \\
    A_\text{near field} &= \frac{\mathcal{P}_\text{abs}}{P_\text{in}}.
    \label{eq:absorptance-nearfield}
\end{align}
\begin{figure}[htpb]
    \centering
    \includegraphics[width=0.9\linewidth]{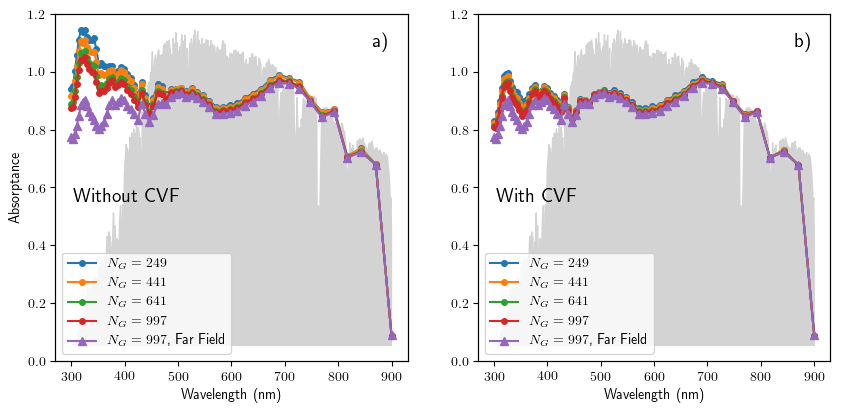}
    \caption{Absorptance calculated from near fields using
    Eqs.~\eqref{eq:power_absorbed_nearfield} - \eqref{eq:absorptance-nearfield}
    (circles). Far field absorptance at $N_G=997$ (triangles). a) S$^4$ implementation
    of RCWA. b) Continuous variable formulation. Gray background shows the AM1.5
    solar spectrum. The CVF shows significant improvement, especially at short
    wavelengths.}
    \label{fig:absorptance-nearfield} 
\end{figure}
The complex dielectric at each frequency is constructed from tabulated real
$n$ and imaginary $k$ parts of the index of refraction in each material
\cite{adachi_optical_1999, hu_optical_2012}.

We calculate $A_\text{near field}$ by extracting $\mathbf{E}(\mathbf{r})$ on a
cubic mesh with 1 nm spacing in the plane for all layers. We use 3 nm spacing
along the z-direction in the ITO layer and 3.5 nm spacing in the nanowire
layer. A sparser mesh of 16 nm spacing is used in the substrate due to the weak
absorption there. This choice of mesh is sufficiently dense to converge the
result better than 1\% using a simple trapezoidal rule integration.  Figure
\ref{fig:absorptance-nearfield}a shows the convergence of $A_\text{near field}$
with $N_G$. Though the near fields are well converged for $\lambda>450$~nm,
they are not converged at short wavelengths even for $N_G=997$.  

In the following sections, we provide two techniques for improving the accuracy
of the near fields in RCWA. The first is an implementation of an existing
technique, which we call the continuous variable formulation (CVF), which
mitigates the Gibbs phenomenon and ensures proper discontinuities at interfaces
by modifying the field computations such that only quantities that are
continuous in real space are reconstructed from their Fourier components. The
discontinuities across in-plane material boundaries are then handled in real
space.  The second technique uses the well-converged, highly accurate far-field
computation of each layer's absorption to rescale the near fields, ensuring
correct total generation within a device layer.

\section{Continuous Variable Formulation}

The CVF is a modification to RCWA that only Fourier reconstructs quantities
which are continuous across material interfaces in real space. These quantities
are the components of the displacement field $\mathbf{D}$ that are normal to,
and the  components of the electric field $\mathbf{E}$ that are tangential to, a
material interface. Using these real-space continuous quantities, one can
determine the full electric field everywhere by using the constitutive
relationship

\begin{equation}
    \mathbf{D} = \epsilon \mathbf{E}
\end{equation}
with a discontinuous real-space $\epsilon$. S$^4$ already uses a related
technique for calculating the Fourier modes, but it does not use this method
when extracting real-space quantities.

To compute the normal and tangential components of any electromagnetic (EM)
field, one must construct a locally-defined vector field that is both tangent
to all material interfaces in the unit cell and periodic in the in-plane
coordinates, which can be generated automatically, as is done by S$^4$
\cite{gotz_normal_2008, liu_s4_2012}.  This vector field induces an associated
projection operator $T$ that can be used to project the Cartesian components of
the EM fields onto this local coordinate system such that:

\begin{align}\label{eq:E_Tdef}
    \begin{bmatrix}
        E_{T,x}(\mathbf{r}) \\
        E_{T,y}(\mathbf{r})
    \end{bmatrix} &= T(\mathbf{r})
    \begin{bmatrix}
        E_{x}(\mathbf{r}) \\
        E_{y}(\mathbf{r})
    \end{bmatrix} \\
    \begin{bmatrix}
        D_{N,x}(\mathbf{r}) \\
        D_{N,y}(\mathbf{r})
    \end{bmatrix} &= N(\mathbf{r})
    \begin{bmatrix}
        D_{x}(\mathbf{r})\\
        D_{y}(\mathbf{r})
    \end{bmatrix} ,
\end{align}
where $N = 1 - T$, $\mathbf{E}_T$ is the component of $\mathbf{E}$ along the
tangential vector field and $\mathbf{D}_N$ is the component of $\mathbf{D}$
perpendicular to the tangential vector field. The total field satisfies

\begin{equation}
    \begin{bmatrix}
        E_{x}(\mathbf{r}) \\
        E_{y}(\mathbf{r})
    \end{bmatrix} =
    \begin{bmatrix}
        E_{T,x}(\mathbf{r}) \\
        E_{T,y}(\mathbf{r})
    \end{bmatrix} +
    \begin{bmatrix}
        E_{N,x}(\mathbf{r}) \\
        E_{N,y}(\mathbf{r})
    \end{bmatrix}.
\end{equation}

By taking the Fourier transform of $T(\mathbf{r})$, the projection onto the tangential vector field
can also be done in Fourier space. Mirroring the notation of Ref.\
\citenum{weismann_accurate_2015}, we denote discrete real-space quantities with
upper case letters (as in $E_x$ to represent the vector $E_x(\mathbf{r}_i)$ for
many points $\mathbf{r}_i$), vectors of Fourier coefficients 
with lower case letters surrounded by single brackets (as in $[e_x]$), and
Fourier space matrix operators with double brackets (as in $\llbracket T
\rrbracket$). Using this notation, the Fourier 
transform of Eq.\ \ref{eq:E_Tdef} is given by \cite{weismann_accurate_2015}:

\begin{align}
    \begin{bmatrix}
    e_{T,x} \\
    e_{T,y}
    \end{bmatrix} &= \llbracket T \rrbracket
    \begin{bmatrix}
    e_{x} \\
    e_{y}
\end{bmatrix}, \label{eq:e_tangential} 
\end{align}
where $\llbracket T \rrbracket$ is the Fourier convolution matrix
\cite{liu_s4_2012, weismann_accurate_2015}. That is, one calculates the Fourier
transform $\tilde{T}(\mathbf{G})$ of $T(\mathbf{r})$, and the ($\mathbf{G}$,
$\mathbf{G'}$)
element of $\llbracket T \rrbracket$ is $\tilde{T}(\mathbf{G}-\mathbf{G'})$. 

We extract $[e_x]$ and $[e_y]$ from a standard RCWA implementation and then
construct $[d_{N,x}]$ and $[d_{N,y}]$. Since $\epsilon(\mathbf{r})$ and
$\mathbf{E}_N(\mathbf{r})$ are both discontinuous, the proper Fourier
factorization takes \cite{li_use_1996}

\begin{align}
    \left \llbracket \frac{1}{\epsilon} \right \rrbracket^{-1} [\mathbf{d}_N]=[\mathbf{e}_N],
\end{align}
where $\llbracket 1/\epsilon \rrbracket^{-1}$ is the $2N_G\times2N_G$ block diagonal 
matrix whose upper-left and lower-right blocks are the inverse of the $N_G\times N_G$
Fourier convolution matrix of $1/\epsilon(\mathbf{r})$. Reference
\citenum{weismann_accurate_2015} showed that the symmetric formulation,

\begin{align}
    \begin{bmatrix}
    d_{N,x} \\
    d_{N,y}
    \end{bmatrix} &=
    \frac{1}{2} \left(
    \llbracket N \rrbracket \left \llbracket \frac{1}{\epsilon} \right \rrbracket^{-1} +
    \left \llbracket \frac{1}{\epsilon} \right \rrbracket^{-1} \llbracket N \rrbracket
    \right)
    \begin{bmatrix}
    e_{x} \\
    e_{y}
    \end{bmatrix}, \label{eq:d_normal}
\end{align}
converges well and conserves power for lossless structures.

\begin{figure}[htpb]
    \centering
    \includegraphics[width=0.8\linewidth]{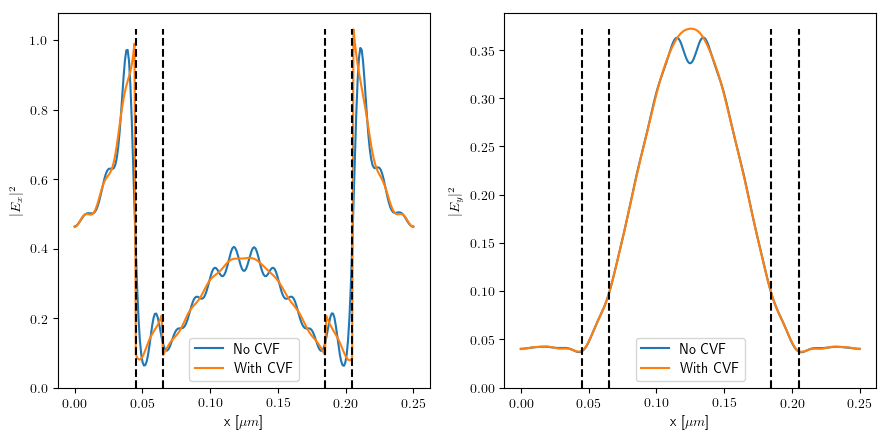}
    \caption{Line cuts of $|E_x|^2$ (left) and $|E_y|^2$ (right) along the
        x-direction through the center of the nanowire 101 nm from the top of
        the nanowire with and without use of the CVF with an incident
        wavelength of 453 nm and $N_G = 997$. The CVF formulation reduces the
        Gibbs oscillations in $E_x$ and introduces proper discontinuities while
        maintaining the continuity of $E_y$.} 
    \label{fig:cvf_nocvf_line_cuts} 
\end{figure}

After finding $d_{N,x}$ and $d_{N,y}$ we reconstruct the real space electric
field

\begin{equation}
    \mathbf{E}_{N}(\mathbf{r}) = \frac{\mathcal{F}^{-1}(\mathbf{d}_N)}{\epsilon_0
    \epsilon_r(\mathbf{r})},
\end{equation}
where $\mathcal{F}^{-1}$ indicates the inverse Fourier transform. This
$\mathbf{E}_N(\mathbf{r})$ has correct discontinuities at material interfaces
where $\epsilon_r$ jumps.  Finally, the real space electric fields in Cartesian
coordinates can be recovered using:

\begin{align}
    E_y(\mathbf{r}) &= \frac{\mathcal{F}^{-1}(d_{N,y})}{\epsilon_0 \epsilon_r(\mathbf{r})} + \mathcal{F}^{-1}(e_{T,y}) \\
    E_x(\mathbf{r}) &= \frac{\mathcal{F}^{-1}(d_{N,x})}{\epsilon_0
    \epsilon_r(\mathbf{r})} + \mathcal{F}^{-1}(e_{T,x}) .
\end{align}
Figure \ref{fig:cvf_nocvf_line_cuts} shows the norm-squared components of the
electric fields computed using unmodified RCWA and the CVF on a line cut along the
x-direction through the center of the nanowire. In this cut, $E_x$ is normal to
the interface and should therefore be discontinuous, while $E_y$ should be
continuous. Note the Gibbs oscillations in the standard result for $E_x$, while
the CVF result has introduced discontinuities at the boundaries and
significantly reduced the amplitude of the Gibbs oscillations.

\begin{figure}[htpb]
    \centering
    \includegraphics[width=0.8\linewidth]{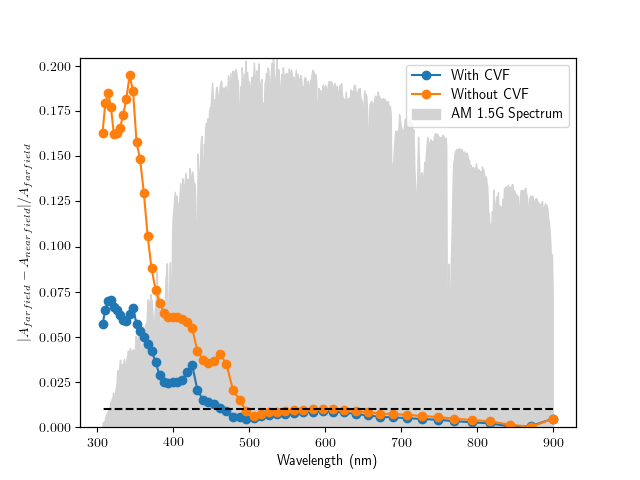}
    \caption{Relative difference between the far field and near field
    calculations of $A$ with $N_G = 997$. Orange line uses the unmodified fields in
    Eq.~\ref{eq:absorptance-nearfield}, blue line uses the CVF fields.  Gray
    background shows the AM1.5G solar spectrum, which is strongest in the region of
    good convergence. Black dashed line indicates the 1\% mark.} 
    \label{fig:absorptance_reldiff} 
\end{figure}

Figure \ref{fig:absorptance-nearfield}b shows the improved agreement between
the CVF absorptance and the well converged far field absorptance. Figure
\ref{fig:absorptance_reldiff} shows the relative difference between the far and
near field absorptances calculated with and without the CVF. It is clear that
the CVF significantly improves the agreement at all wavelengths, but the
disagreement is still significant for wavelengths shorter than 450 nm. Figure
\ref{fig:absorptance_reldiff} indicates the AM1.5G spectrum, which shows that
the CVF-based $A_\text{near field}$ agrees well with the far field results
through the most important parts of the solar spectrum. In the next section, we
introduce a simple rescaling technique to increase accuracy of the near fields
at all incident wavelengths.

\section{Rescaling Technique}

In device simulations, the total optical generation rate must be determined
accurately. The exact position where generation occurs is somewhat less
important, as the carriers drift and diffuse, and deviations on the scale of a
few nanometers are rarely significant.  We can ensure that the total generation
in each layer is calculated correctly, even with inexpensive RCWA calculations
that have not fully converged the local fields. To achieve this goal, we use
the well-converged far field results (as shown in Fig.
\ref{fig:absorptance-farfield}) to rescale the components of the near fields in
each layer such that $A_\text{far field}$ and $A_\text{near field}$ agree exactly.
This rescaling can be done by defining a rescaling factor $F$ for each layer
$i$ and frequency $\omega$: 

\begin{equation}
    F_i(\omega) = \frac{A_\text{far field}^i(\omega)}{A_\text{near field}^i(\omega)}.
\end{equation}
Then, the components of $\mathbf{E}$ may be rescaled such that:

\begin{equation}
    \mathbf{E}^\text{rescaled}(\omega) = \sqrt{F_i(\omega)} \mathbf{E}(\omega)
\end{equation}
for fields in the appropriate layer. 

\begin{figure}[htpb]
    \centering
    \includegraphics[width=0.8\linewidth]{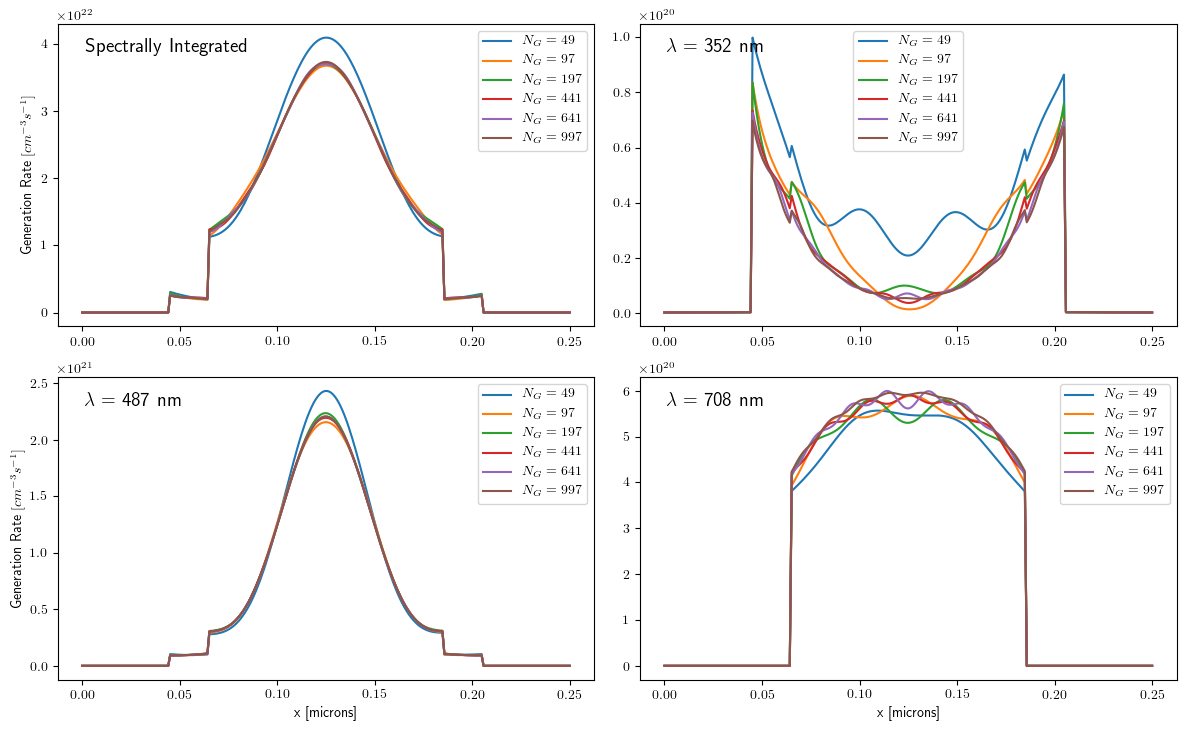}
    \caption{Rescaled generation rate on a line cut along the x direction
    through the center of the nanowire 83 nm from the top of the nanowire
    layer. The spectrally integrated and $\lambda = 487$ nm case are clearly
    converged even at $N_G = 197$, while the longer and shorter wavelengths need
    high $N_G$ to remove all the Gibbs oscillations.} 
    \label{fig:rescaled_generation_line_cuts}
\end{figure}

\begin{figure}[htpb]
    \centering
    \includegraphics[width=0.8\linewidth]{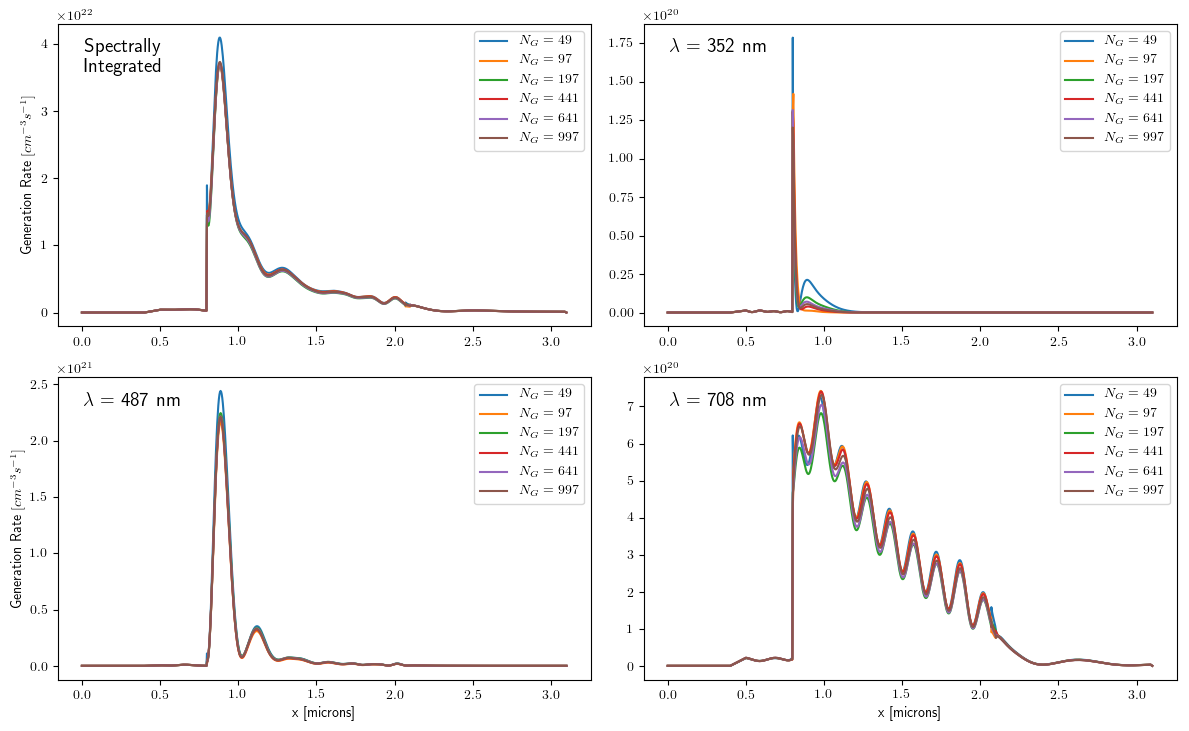}
    \caption{Rescaled generation rate on a line cut along the z direction
    through the center of the nanowire core.}
    \label{fig:rescaled_generation_line_cuts_alongz}
\end{figure}

\begin{figure}[htpb]
    \centering
    \includegraphics[width=0.8\linewidth]{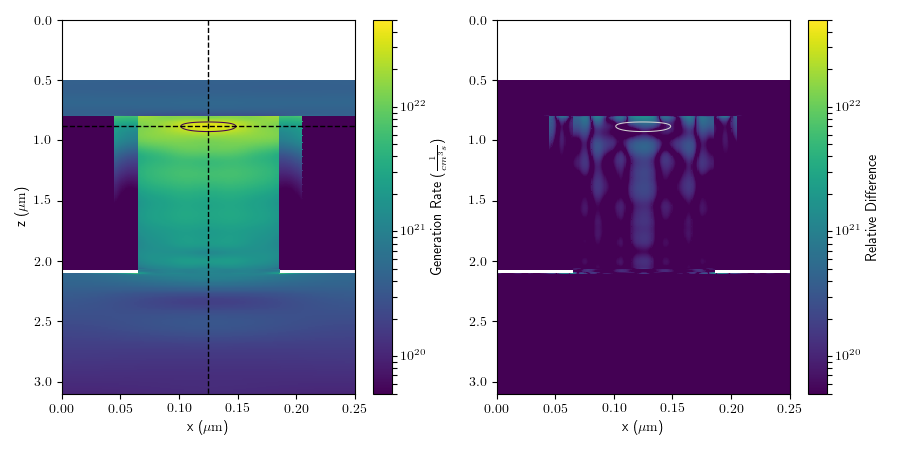}
    \caption{Left: Spectrally integrated generation rate with AM1.5G spectrum
        along a cut through the middle of the nanowire using the rescaled
        fields with $N_G = 197$.  Right: Absolute difference between the
        generation rate shown at left and the well-converged, rescaled
        generation rate at $N_G = 997$. The deviations between the two
        generation maps are small.  White regions are areas of vacuum and
        SiO$_2$, where the generation rate is zero. Area within the solid contour indicates
        the location of peak generation, greater than $2.75 \times 10^{22}
        \text{ cm}^{-3} \text{s}^{-1}$, while the differences there are much smaller.
        Dashed lines indicate line cuts shown in Figs.\
        \ref{fig:rescaled_generation_line_cuts} and
        \ref{fig:rescaled_generation_line_cuts_alongz}.}
    \label{fig:rescaled_generation_rate} 
\end{figure}

This rescaling technique allows accurate determination of spectrally-integrated
generation rates with small numbers of basis terms. Figures
\ref{fig:rescaled_generation_line_cuts} and
\ref{fig:rescaled_generation_line_cuts_alongz} show line cuts through the test
structure at three representative wavelengths and spectrally integrated under
AM1.5G illumination \cite{noauthor_standard_2012}, calculated with 60 equally
spaced frequencies. At 487 nm, the fields are quantitatively converged at small
$N_G$, while the Gibbs oscillations are not entirely removed either at shorter
or longer wavelengths.  Calculations dependent on spectrally resolved local
fields, such as external quantum efficiency (EQE), thus require relatively
large $N_G$ at some wavelengths. When the fields are spectrally integrated,
however, the essentially random phases of the oscillations average away, and
the spectrally-integrated fields are quantitatively converged by $N_G=197$.
Figure \ref{fig:rescaled_generation_rate} compares the rescaled
spectrally-integrated generation rates along a plane through the center of the
nanowire at $N_G = 997$ and 197, showing the excellent agreement that
rescaling permits, even at low $N_G$. This spectrally integrated generation
rate is sufficient for optoelectronic modeling while reducing the requirements
for $N_G$ by a factor of 5.

\begin{figure}[htpb]
    \centering
    \includegraphics[width=0.8\linewidth]{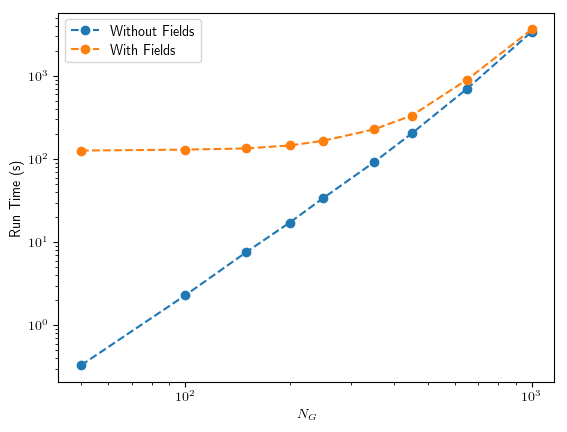}
    \caption{Run time of a single simulation as a function of basis terms with
    and without computation of the local fields. Field computations dominate
    the runtime at small $N_G$. A least-squares fit to the blue line
    yields a slope of 3.06, consistent with the $N_G^3$ scaling of the QR algorithm
    for solving eigenvalue problems. Dashed lines are guides to the eye.}
    \label{fig:runtimes}
\end{figure}

The computational cost of the RCWA method scales as $N_G^3$, so reducing $N_G$
by a factor of 5 (from 1000 to 200) theoretically reduces the runtime by a factor of 125,
and reducing to $N_G=100$ can reduce the runtime by a factor of 1000.
Extracting the electric fields on a dense mesh of points, however, also has a
computational cost, and for sufficiently small $N_G$, this electric-field
extraction limits the runtime. Figure \ref{fig:runtimes} shows an estimate of
the simulation run times for a single incident wavelength. Each desired
wavelength must be calculated separately, and they all take approximately the
same amount of processor time. Simulations were run on a single core of an
Intel Xeon E5-2640 v4 CPU with a 2.40GHz clock speed. The figure shows both the
simulation time for RCWA to determine the field amplitudes, in Fourier space,
and for the extraction of those fields on the dense real-space mesh described
above. The CVF method does not significantly change the run times. Efforts were
made to minimize data input/output time and resource contention in all
benchmarks.  These results show that running at $N_G=197$ has a cost 25 times
less than at $N_G=997$, and that cost is dominated by the field calculation and
export. The field calculation and export time can possibly be optimized further
and would certainly be reduced if a coarser mesh were requested. Decreasing %
that cost could allow computation times to be reduced by an additional factor
of 5. Simulations at each frequency are completely independent, so      %
computation time for a full spectral sweep can be easily reduced by running all
frequencies in parallel \cite{tange_ole_2018_1146014}. With a sufficient number
of CPU cores, a full spectral sweep can be run in the same clock time as a
single simulation.

\section{Conclusion}

In this work, we investigate the accuracy of RCWA for optical modeling of
nanowire solar cells. We find excellent accuracy with low computational
cost at long incident wavelengths, but poor accuracy at short incident
wavelengths. To increase the accuracy of RCWA we extend the open-source
library S$^4$ \cite{liu_s4_2012} to include an already published technique for
improving near field computations in RCWA \cite{weismann_accurate_2015}. Our
implementation mitigates the Gibbs phenomenon and introduces physically
expected discontinuities in the fields at material interfaces, improving
convergence of the near fields at all incident wavelengths. To bring
convergence within a 1\% tolerance, we introduce a simple rescaling technique
that uses the well converged far field quantities to rescale the near fields on
a per layer basis. These improvements open up the possibility of using RCWA as
a low cost optical modeling technique in a full optoelectronic device model of
nanowire solar cells.

\section{Acknowledgements}

We thank Anna Trojnar for helpful conversations and acknowledge funding from
the Natural Sciences and Engineering Research Council of Canada.

%\bibliography{main}

\end{document}